# Spatiotemporal self-similar fiber laser

Uğur Teğin,[1,2,*] Eirini Kakkava,[1] Babak Rahmani,[2] Demetri Psaltis,[1] and Christophe Moser[2]

[1]Optics Laboratory, École Polytechnique Fédérale de Lausanne, Lausanne, Switzerland
[2]Laboratory of Applied Photonics Devices, École Polytechnique Fédérale de Lausanne, Lausanne, Switzerland
*ugur.tegin@epfl.ch

**Abstract:** In this Letter, we demonstrate, to the best of our knowledge, the first spatiotemporally mode-locked fiber laser with self-similar pulse evolution. The multimode fiber oscillator generates parabolic amplifier similaritons at 1030 nm with 90 mW average power, 2.3 ps duration, and 37.9 MHz repetition rate. Remarkably, we observe experimentally a near-Gaussian beam quality ($M^2<1.4$) at the output of the highly multimode fiber. The output pulses are compressed to 192 fs via an external grating compressor. Numerical simulations are performed to investigate the cavity dynamics which confirm experimental observations of self-similar pulse propagation. The reported results open a new direction to investigate new types of pulse besides beam shaping and nonlinear dynamics in spatiotemporal mode-locked fiber lasers.

For single-mode fiber lasers, ytterbium-based laser systems are generally preferred to achieve high power pulses with relatively broadband spectra. Over the years, dispersion engineering techniques have been demonstrated to obtain various ultrashort pulse types in ytterbium-based fiber cavities such as soliton [1], dispersion-managed [2] and passive self-similar pulses [3]. In 2006, Chong et al. demonstrated the first dissipative soliton pulse formation with an all-normal-dispersion cavity [4]. For ytterbium-based laser systems, all-normal dispersion cavities provide a simple platform to build all-fiber dissipative soliton lasers [5,6]. The generation of mode-locked dissipative soliton pulses is attributed to spectral intracavity filtering by using 8-12 nm bandpass filters. Interestingly, when the intra-cavity bandpass filter becomes narrow (≤4 nm), the temporal pulse shape changes from Gaussian to parabolic [7]. These parabolic pulses are referred to as self-similar (amplifier similariton) pulses. Initially, self-similar pulses were studied and proposed in fiber amplifiers [8,9] as an alternative to chirped pulse amplification [10]. The similariton pulses have a parabolic pulse shape and they are a class of solution to the nonlinear Scrödinger equation with a gain term [8]. Renninger et al. reported that, in an all-normal-dispersion cavity design, with strong spectral filtering (≤4 nm), an amplifier similariton pulse can be generated in the gain segment of the laser cavity. Compared to dissipative soliton pulses the amplifier similaritons experience large spectral breathing (the ratio of spectral bandwidth before and after the spectral filter is >5). This distinctive behavior generates mode-locked pulses with a broader spectrum and hence potentially shorter pulse duration. Because the chirp of the output pulse depends solely on the gain segment, the amplifier similariton pulses also feature less chirp than given by the total cavity dispersion [7]. Recently, Ma et al. demonstrated a ytterbium-based all-normal self-similar mode-locked fiber laser tunable from 1030 nm to 1100 nm by suppressing amplified spontaneous emission by heating the gain fiber [11]. All these studies were done in single-mode laser cavities.

In recent years, graded index multimode fibers (GIMFs) have become subject to extensive study due to their unique nonlinear properties and potential higher power handling capacities. With low modal dispersion and periodic self-imaging, spatiotemporal pulse propagation of high power pulses in GIMF generates interesting nonlinear effects such as spatiotemporal instability [12,13], dispersive wave generation [14], graded-index solitons [15,16], self-beam cleaning [17], nonlinear pulse compression [18] and supercontinuum generation [19,20]. In addition to the aforementioned studies, spatiotemporal mode-locking was demonstrated with graded-index multimode fibers in multimode fiber laser cavities by Wright et al. [21]. Later, observations of the bound state solitons (soliton molecules) and harmonic mode-locking reported in spatiotemporal mode-locked fiber lasers [22,23]. All of these studies presented laser cavities featuring dissipative soliton pulses with Gaussian temporal profile and low output beam quality. However, to the best of our knowledge, there are no reports on the generation of different temporal pulse types than dissipative solitons in spatiotemporal mode-locked lasers.

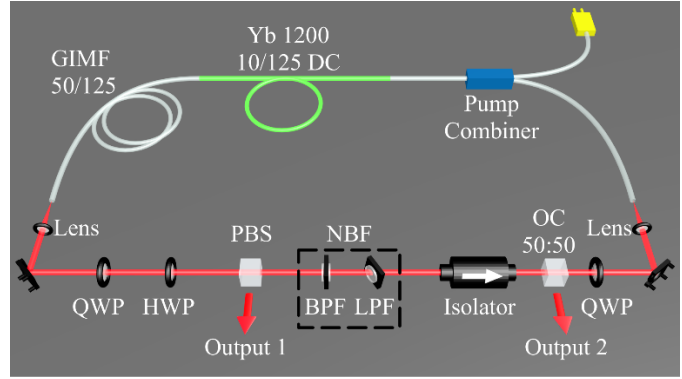

**Fig. 1.** Schematic of the spatiotemporal self-similar fiber laser: QWP, quarter-wave plate; HWP, half-wave plate; PBS, Polarizing Beamsplitter; NBF, narrow bandpass filter; BPF, bandpass filter; LPF, long-pass filter; OC, output coupler; GIMF, graded-index multimode fiber.

Here, we present the first spatiotemporal self-similar fiber laser capable of generating amplifier similariton pulses with a parabolic temporal shape. Interestingly, the train of the mode-locked pulse coming out of the 250 mode GIMF has a stable good beam quality of $M^2$-value <1.4. The presented laser is an all-normal-dispersion cavity containing a GIMF with 50 $\mu$m core diameter, a step-index multimode gain and passive fiber segment both with a 10 $\mu$m core diameter. Mode-locking is achieved by nonlinear polarization evolution (NPE) [24]. A narrow bandpass spectral filter with 3.8 nm bandwidth is constructed with a cascade of a bandpass filter with 10 nm bandwidth and a tilted longpass filter. First, numerical simulations are performed to investigate the possibility of amplifier similariton formation in the multimode laser cavity. Encouraged by the numerical simulations, experimental studies were performed with numerically obtained cavity parameters such as fiber length of each cavity segment. We experimentally achieved self-starting mode-locking for which the laser generates amplifier similaritons at 1030 nm with 90 mW average power, 2.4 nJ pulse energy and ~38 MHz repetition rate. The experimental results indicate that the pulse experiences 6-fold increase in its spectral width inside the laser cavity. The chirped output pulse duration is measured as 2.3 ps which is remarkably short when compared with the group-velocity dispersion of the laser cavity. Both of these features point to an amplifier similariton behavior. The chirped output pulses are externally compressed to 240 fs by a grating pair as measured by a second-order autocorrelation. By reconstructing the pulse profile using the phase and intensity from cross-correlation and spectrum only (PICASO) method [25], we obtained a 172 fs pulse duration with a parabolic temporal shape.

The schematic of the fiber laser is illustrated in Fig. 1. The numerical simulations are performed based on this cavity design to investigate the possibility of self-similar pulse formation. We used the fourth-order Runge-Kutta in the interaction picture method to simulate the pulse propagation [26]. A Gaussian intracavity bandpass filter is assumed with 4 nm spectral bandwidth. We modeled the GIMF segment as a periodically oscillating cubic nonlinearity term to reach manageable computation time, thus discarding the spatial information about the pulse propagation [16,27]. For the GIMF segment, the maximum fiber index contrast is 0.01, the fiber

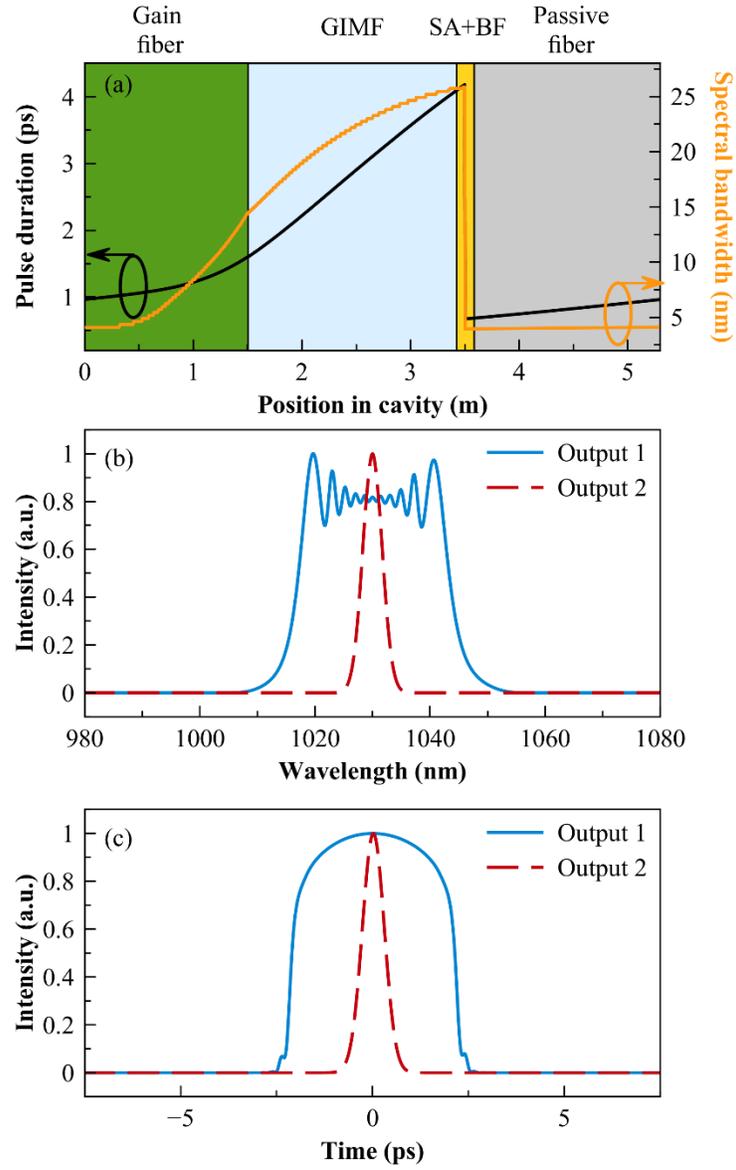

**Fig. 2.** (a) Simulated pulse duration and spectral bandwidth variation over the cavity: SA, saturable absorber; BF, bandpass filter. (b) Simulated laser spectra obtained from the defined output couplers. (c) Simulated temporal profile obtained at output 1 and output 2.

core index is 1.444 and the input beam FWHM is set at 10 μm. The gain coefficient is defined as

$$g(E_p) = \frac{g_0}{1 + E_p/E_{sat}}$$

where $E_p$ is the pulse energy, $g_0$ is the small-signal gain and $E_{sat}$ is the gain saturation energy. The gain is modeled as Lorentzian with 50 nm bandwidth, 30 dB small-signal gain and 1.35 nJ saturation energy. In the numerical studies, the silica fiber group-velocity dispersion ($\beta_2$) is set to 24.8 $fs^2/m$ for passive and 26.16 $fs^2/m$ for the active fiber. The nonlinearity coefficients ($\gamma$) for the passive and active fiber are 2.12 $(kWm)^{-1}$ and 4.14 $(kWm)^{-1}$, respectively. Raman scattering and shock terms are also incorporated in our numerical studies. We seed the simulations with quantum noise fields. A stable mode-lock regime is achieved after 25 round-trips as illustrated in Fig. 2. The spectral and temporal evolution of the mode-locked pulse inside the cavity is shown in Fig. 2(a). In the gain fiber, substantial spectral broadening is observed and after the GIMF segment, the spectral bandwidth of the pulse reaches ~26 nm. This broad spectrum is reduced to 4 nm with a bandpass filter. Thus, the pulse experiences 6.5 times spectral broadening in one round-trip (Fig. 2(b)). The temporal profile of the pulse is presented in Fig. 2(c) and at the output 1 (before filter) and output 2 (after filter), a pulse duration of 4.2 ps and 650 fs is achieved respectively. Output 1 and output 2 are the NPE output and the 50:50 output couplers, respectively. For numerical simulations, 2.5 nJ pulse energy is observed at the output 1.

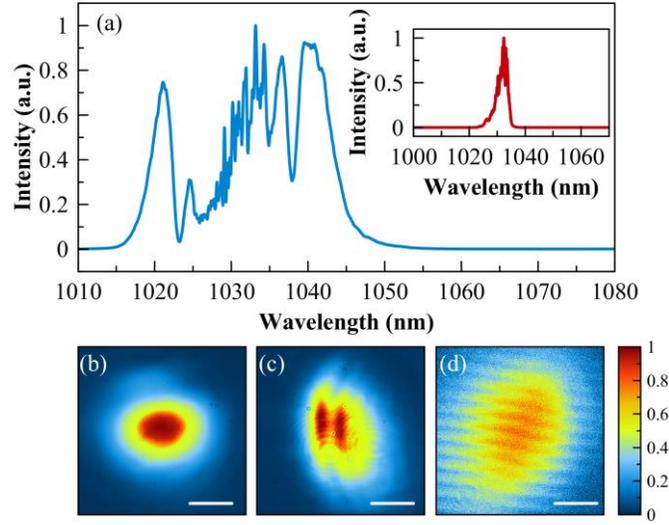

**Fig. 3.** (a) Measured spectrum from output 1 and output 2 (inset). Measured beam profiles case from output 1 (b) and output 2 (c) for mode-locked operation. (d) Measured beam profile from output 1 for continuous-wave operation case. Scale bars indicated in beam profiles are 520 μm.

The experiments were then performed with the fiber lengths obtained numerically. A pump combiner with 10 μm core diameter is integrated to the cavity to couple the 976 nm high power fiber-coupled diode laser for pumping the 1.5 m highly doped ytterbium fiber (nLight Yb-1200-10/125) gain segment. The fiber sections with 10 μm core diameter support 3 modes at 1 μm. This passive fiber section with 10 μm core diameter is 1.8 m long. In order to excite the higher order modes of the 250 mode GIMF (Thorlabs GIF50C), the gain fiber is spliced to the 2 m GIMF with a small off-set (~5 μm). Mode-locking was achieved by adjusting the intracavity wave plates. At approximately 1.5 W pump power, self-starting mode-locking with a repetition rate of ~38 MHz is observed. Experimentally obtained optical spectra and beam profiles from the output couplers before and after the bandpass filter are presented in Fig. 3. A drastic improve is observed in output beam profile when the laser operation changed from continuous-wave to mode-locked (see Fig. 3(b) and Fig. 3(d)). The spectral width of the amplifier similariton pulses reach to 24 nm after the GIMF and the spectral profile is reshaped to 3.8 nm with the narrow bandpass filter. This result is consistent with the pulse breathing ratio of >5 found after one round-trip in a single-mode amplifier similariton laser [7]. The measured output power of the laser is 90 mW and 10 mW at the output 1 and output 2, respectively. After the narrow bandpass intracavity spectral filter, the pulses experience spatial filtering due to the aperture of the isolator (see Fig. 3(a-inset)). We observe that spatial filtering is necessary to achieve spatiotemporal mode-locking similarly to dissipative soliton pulses in multimode fiber lasers [21,22]. Beam profiles are measured by a 4f-system with 1.5 magnification. As shown in Fig. 3(a), the beam at the output port 1, which is immediately after the 250 mode GIMF, has a near symmetric shape. $M^2$-measurements are performed to determine the quality of this beam and presented in Fig. 4(a). When we compare with the spatiotemporal dissipative soliton lasers, the self-similar spatiotemporal laser produces a better beam quality. We believe a possible explanation is that the phase of each of the multiple fiber modes are such that their interference produces a near-Gaussian beam similarly to what was demonstrated for single-pass beam-shaping with multimode fibers [28,29]. This particular phase condition may automatically be selected by the laser cavity to produce a beam which has smaller loss after passing through the spatial filter and coupled to the 10 $\mu$m core fiber. We numerically investigate the single-pass pulse propagation inside a 2m GIMF by considering the interactions between the first 7 LP (m=0) with coupled-mode analysis for a pump pulse with 2.3 ps duration and ~8nJ energy [15,30]. In these studies, we notice that energy transfer to lower order modes occurs for various initial conditions. The observed beam-shaping effect needs to be further investigated in future work.

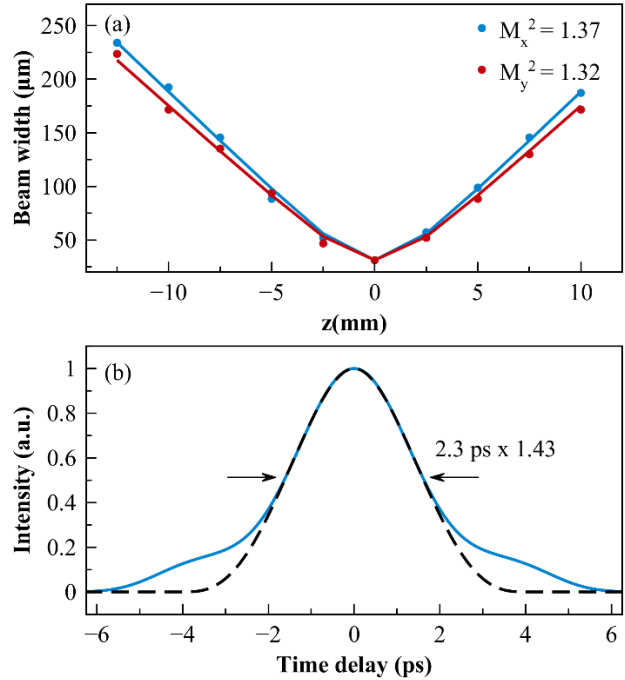

**Fig. 4.** (a) $M^2$-measurement of the beam from output 1. (b) Autocorrelation trace of the chirped pulse measured from output 1 (solid) and theoretical fit with chirp-free parabolic pulse shape (dashed).

For the pulses from output 1, no secondary pulse formation or periodic oscillation of the pulse train was observed. The duration of the chirped output pulses are 2.3 ps with a 1.43 deconvolution factor as shown in Fig. 4(b). The chirped pulse duration is remarkably smaller for an all-normal-dispersion cavity with total cavity dispersion ∼0.13 ps$^2$ and it validates the self-similar behavior of the pulses inside the laser cavity. These pulses are dechirped (compressed) using an external grating compressor with a 300 line/mm diffraction grating pair to 192 fs (Fig. 5(a)). We utilize the PICASO algorithm to retrieve the temporal profile from the spectrum and autocorrelation data. As presented in Fig. 5(b), the resulting pulse features 172 fs pulses duration with a parabolic temporal profile.

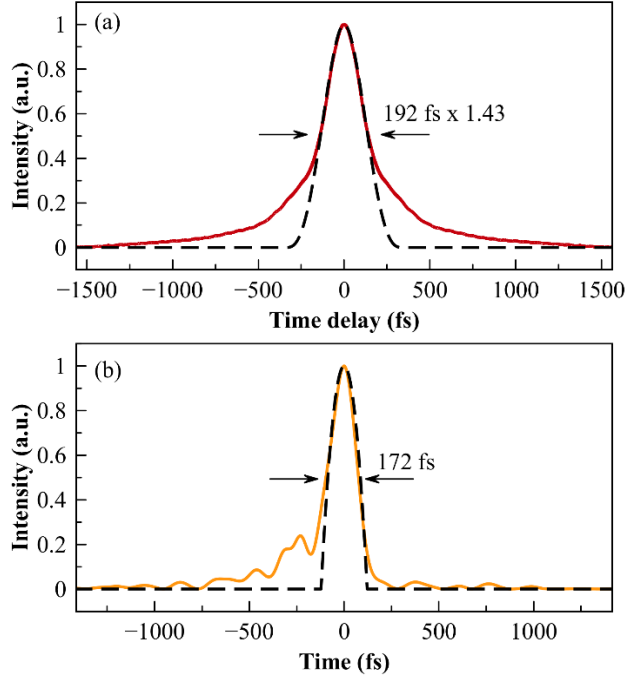

**Fig. 5.** (a) Autocorrelation trace of the compressed pulse measured from output 1 and theoretical fit with chirp-free parabolic pulse shape (dashed). (b) PICASO retrieved dechirped pulse shape and parabolic pulse shape fit (dashed).

We test the power handling capacity of the spatiotemporal self-similar fiber laser, by gradually increasing the pump power. Up to 195 mW average power from output port 1, a mode-locked operation with a single pulse on a cavity round-trip, was maintained. This average power was achieved with 2.1 W pump power level. Pulses were yielding 5.1 nJ pulse energy. A broader spectrum (>35 nm) is obtained but the temporal pulse shape is degraded and the compressed pulse duration is increased (>300 fs).

In conclusion, we numerically and experimentally demonstrate an all-normal-dispersion ytterbium-based spatiotemporal mode-locked fiber laser supporting self-similar pulse evolution, for the first time in the literature. The laser dynamics is numerically and experimentally validated. We obtained intracavity large spectral breathing (>6) and low chirp of the output pulses. Our observations verified the spatiotemporal self-similar pulse formation and the parabolic pulse generation. The oscillator generates amplifier similariton pulses at 1030 nm with 90 mW average power, 2.4 nJ energy, 2.3 ps duration and ∼38 MHz repetition rate. Pulses are dechirped to 192 fs via an external grating compressor. Contrary to the beam profile obtained in dissipative multimode mode-locked cavities, we measured a near-Gaussian beam profile of the amplifier similariton pulses with M$^2$-value <1.4. The mechanism for the generation of the good beam quality from the cavity is not fully understood and will be the subject of further research. Nonetheless, the combination of good beam quality, sub-200 fs self-similar pulse from a multimode cavity is a promising platform to generate high power ultrashort pulses. We believe the reported observations are of great interest for nonlinear pulse propagation, pulse and beam shaping in spatiotemporal mode-locked fiber lasers. The presented technique can find applications in wavelength and pulse shape tunable laser sources.

**Acknowledgments**

The authors thank Prof. C. Brès and PHOSL for lending the fiber splicer used in the experiments.

**References**

1. Isomäki, A., & Okhotnikov, O. G. (2006). Femtosecond soliton mode-locked laser based on ytterbium-doped photonic bandgap fiber. Optics express, 14(20), 9238-9243.
2. Ortaç, B., Hideur, A., Chartier, T., Brunel, M., Özkul, C., & Sanchez, F. (2003). 90-fs stretched-pulse ytterbium-doped double-clad fiber laser. Optics letters, 28(15), 1305-1307.
3. Ilday, F. Ö., Buckley, J. R., Clark, W. G., & Wise, F. W. (2004). Self-similar evolution of parabolic pulses in a laser. Physical review letters, 92(21), 213902.
4. Chong, A., Buckley, J., Renninger, W., & Wise, F. (2006). All-normal-dispersion femtosecond fiber laser. Optics express, 14(21), 10095-10100.
5. Özgören, K., & Ilday, F. Ö. (2010). All-fiber all-normal dispersion laser with a fiber-based Lyot filter. Optics letters, 35(8), 1296-1298.
6. Teğin, U., & Ortaç, B. (2018). All-fiber all-normal-dispersion femtosecond laser with a nonlinear multimodal interference-based saturable absorber. Optics letters, 43(7), 1611-1614.
7. Renninger, W. H., Chong, A., & Wise, F. W. (2010). Self-similar pulse evolution in an all-normal-dispersion laser. Physical Review A, 82(2), 021805.
8. Fermann, M. E., Kruglov, V. I., Thomsen, B. C., Dudley, J. M., & Harvey, J. D. (2000). Self-similar propagation and amplification of parabolic pulses in optical fibers. Physical Review Letters, 84(26), 6010.
9. Kruglov, V. I., Peacock, A. C., Dudley, J. M., & Harvey, J. D. (2000). Self-similar propagation of high-power parabolic pulses in optical fiber amplifiers. Optics letters, 25(24), 1753-1755.
10. Strickland, D., & Mourou, G. (1985). Compression of amplified chirped optical pulses. Optics communications, 55(6), 447-449.
11. Ma, C., Khanolkar, A., & Chong, A. (2019). High-performance tunable, self-similar fiber laser. Optics letters, 44(5), 1234-1236.
12. Krupa, K., Tonello, A., Barthélémy, A., Couderc, V., Shalaby, B. M., Bendahmane, A., Millot, G., & Wabnitz, S. (2016). Observation of geometric parametric instability induced by the periodic spatial self-imaging of multimode waves. Physical review letters, 116(18), 183901.
13. Teğin, U., & Ortaç, B. (2017). Spatiotemporal instability of femtosecond pulses in graded-index multimode fibers. IEEE Photonics Technology Letters, 29(24), 2195-2198.
14. Wright, L. G., Christodoulides, D. N., & Wise, F. W. (2015). Controllable spatiotemporal nonlinear effects in multimode fibres. Nature Photonics, 9(5), 306.
15. Renninger, W. H., & Wise, F. W. (2013). Optical solitons in graded-index multimode fibres. Nature communications, 4, 1719.
16. Ahsan, A. S., & Agrawal, G. P. (2018). Graded-index solitons in multimode fibers. Optics letters, 43(14), 3345-3348.
17. Krupa, K., Tonello, A., Shalaby, B. M., Fabert, M., Barthélémy, A., Millot, G., Wabnitz, S., & Couderc, V. (2017). Spatial beam self-cleaning in multimode fibres. Nature Photonics, 11(4), 237.
18. Krupa, K., Tonello, A., Couderc, V., Barthélémy, A., Millot, G., Modotto, D., & Wabnitz, S. (2018). Spatiotemporal light-beam compression from nonlinear mode coupling. Physical Review A, 97(4), 043836.
19. Lopez-Galmiche, G., Eznaveh, Z. S., Eftekhar, M. A., Lopez, J. A., Wright, L. G., Wise, F., Christodoulides, D. & Correa, R. A. (2016). Visible supercontinuum generation in a graded index multimode fiber pumped at 1064 nm. Optics letters, 41(11), 2553-2556.
20. Teğin, U., & Ortaç, B. (2018). Cascaded Raman scattering based high power octave-spanning supercontinuum generation in graded-index multimode fibers. Scientific reports, 8(1), 12470.
21. Wright, L. G., Christodoulides, D. N., & Wise, F. W. (2017). Spatiotemporal mode-locking in multimode fiber lasers. Science, 358(6359), 94-97.
22. Qin, H., Xiao, X., Wang, P., & Yang, C. (2018). Observation of soliton molecules in a spatiotemporal mode-locked multimode fiber laser. Optics letters, 43(9), 1982-1985.
23. Ding, Y., Xiao, X., Wang, P., & Yang, C. (2019). Multiple-soliton in spatiotemporal mode-locked multimode fiber lasers. Optics Express, 27(8), 11435-11446.

24. Hofer, M., Ober, M. H., Haberl, F., & Fermann, M. E. (1992). Characterization of ultrashort pulse formation in passively mode-locked fiber lasers. IEEE journal of quantum electronics, 28(3), 720-728.
25. Nicholson, J. W., Jasapara, J., Rudolph, W., Omenetto, F. G., & Taylor, A. J. (1999). Full-field characterization of femtosecond pulses by spectrum and cross-correlation measurements. Optics letters, 24(23), 1774-1776.
26. Hult, J. (2007). A fourth-order Runge–Kutta in the interaction picture method for simulating supercontinuum generation in optical fibers. Journal of Lightwave Technology, 25(12), 3770-3775.
27. Conforti, M., Arabi, C. M., Mussot, A., & Kudlinski, A. (2017). Fast and accurate modeling of nonlinear pulse propagation in graded-index multimode fibers. Optics letters, 42(19), 4004-4007.
28. Morales-Delgado, E. E., Farahi, S., Papadopoulos, I. N., Psaltis, D., & Moser, C. (2015). Delivery of focused short pulses through a multimode fiber. Optics express, 23(7), 9109-9120.
29. Morales-Delgado, E. E., Psaltis, D., & Moser, C. (2015). Two-photon imaging through a multimode fiber. Optics express, 23(25), 32158-32170.
30. Mafi, A. (2012). Pulse propagation in a short nonlinear graded-index multimode optical fiber. Journal of Lightwave Technology, 30(17), 2803-2811.